\documentclass{ws-procs9x6}
\newcommand{\be}{\begin{eqnarray}}
\newcommand{\ee}{\end{eqnarray}}
\newcommand{\tr}{\rm Tr}
\newcommand{\hmu}{\hat{\mu}}
\newcommand{\hm}{\hat{m}}
\newcommand{\hx}{\hat{x}}
\newcommand{\hy}{\hat{y}}
\newcommand{\htt}{\hat{t}}
\newcommand{\trho}{\tilde{\rho}}
\def\conj#1{{{#1}^{*}}}

\newcommand\nn{\nonumber}

\begin{document}

\title{Statistical QCD with non-positive measure}

\author{J.C. Osborn}

\address{Argonne Leadership Computing Facility\\ 9700 S. Cass Avenue,
Argonne, IL 60439, USA}
\address{Center for Computational Science, Boston University \\
Boston, MA 02215, USA}

\author{K. Splittorff}

\address{The Niels Bohr Institute\\
Blegdamsvej 17, DK-2100,
Copenhagen {\O}, Denmark \\
{\sl speaker}, $^*$E-mail: split@nbi.dk}

\author{J.J.M.\ Verbaarschot}

\address{Department of Physics and Astronomy, SUNY \\ Stony Brook,
New York 11794, USA \\
E-mail: verbaarschot@cs.physics.sunysb.edu}

\begin{abstract}
In this talk we discuss the microscopic limit of QCD at nonzero chemical
potential. In this domain, where the QCD partition function is under complete
analytical control, 
we uncover an entirely new 
link between the spectral density of the Dirac operator and the chiral
condensate: violent complex oscillations on the microscopic scale give rise to 
 the
discontinuity of the chiral condensate at zero quark mass. We
first establish this relation exactly within a random matrix framework and
then analyze the importance of the individual modes by Fourier analysis.

\end{abstract}

\keywords{nonperturbative QCD, sign problem, random matrix theory}

\bodymatter

\section{Introduction}
\label{sec:intro}

Many firm results about QCD are based on the fact that the Euclidean partition
function has a real and positive measure. Prime examples are QCD
inequalities and the evaluation of the  
lattice QCD partition function by Monte Carlo simulations. 
For this reason much of the intuition
gained about nonperturbative QCD is based on a probabilistic
interpretation. If we introduce a chemical potential in order to favor the
presence of quarks over anti-quarks the measure in the Euclidean partition
function is no longer positive definite. In this lecture we will analyze 
the link
between the spectral density of the QCD Dirac operator and the order parameter
for spontaneous chiral symmetry breaking. We will see that the intuition
based on zero chemical potential results fails completely at nonzero chemical
potential. 

At zero chemical potential the eigenvalues of the anti-Hermitean
Dirac operator are located
on the imaginary axis. The chiral
condensate can be thought of as the electric field at the position on the real
axis given by the quark mass, $m$, created by positive charges located at the  
purely imaginary eigenvalues \cite{Barbour}. The chiral condensate is 
the jump in the
electric field as the quark mass passes through zero. Clearly such a
jump will occur if the density of the  eigenvalues/charges  near the origin of
the imaginary axis scales with the space-time volume \cite{BC}. 

\begin{figure}
\begin{center}
\psfig{file=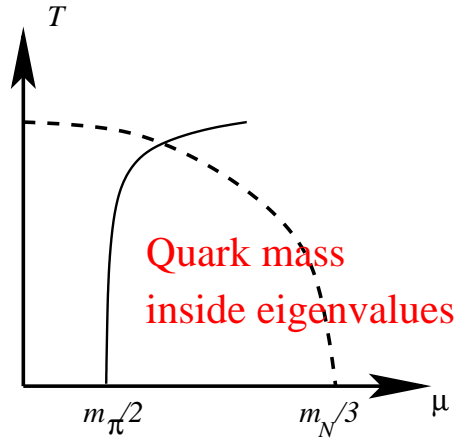,width=6cm}
\end{center}
\caption{Schematic phase diagram of QCD and region where the quark mass is
  inside the support of the eigenvalue density. The dashed curve indicates the
  chiral transition be it first order or crossover. To the right of the full
  line the quark mass is inside the support of the Dirac spectrum. The 
solid curve
  does {\sl not} indicate a phase transition in QCD.}
\label{fig:PD}
\end{figure}

At nonzero chemical potential the anti-Hermiticity of the Dirac operator is
lost and the eigenvalues are located homogeneously in a strip parallel to the
imaginary axis with width proportional to $\mu^2$ (for sufficiently large
 $\mu$). The constant of
proportionality is such that, at sufficiently low temperatures, 
the quark mass hits the eigenvalue strip for $\mu=m_\pi/2$ 
\cite{Gibbs,TV,LKS}. From our intuition
at zero chemical potential we are lead to the conclusion 
that at any (however small) nonzero value of the chemical potential 
the chiral condensate has no discontinuity when
the quark mass crosses the imaginary axis. 
 This is in sharp contrast to the
expected phase diagram, see ~Fig. \ref{fig:PD}. 
(Notice however, that in phase quenched QCD 
the chiral condensate rotates into a pion condensate so that chiral
symmetry remains broken spontaneously in spite of a vanishing chiral 
condensate.) Increasing the temperature
tends to decrease the width of the strip of 
eigenvalues so that the value of the chemical potential for which the quark
mass is inside the domain of the eigenvalues increases as well (see Fig.
\ref{fig:PD} for illustration).  

At nonzero chemical potential the eigenvalue density of the Dirac operator is defined as 
\be
\rho_{N_f}(x,y) = 
\left\langle \sum_k \delta(x-x_k)\delta(y-y_k) \right\rangle_{N_f}.
\label{rhoDef}
\ee
Since the measure includes the fermion determinants there is no guarantee
that the density is real or positive. To completely destroy
the electrostatic picture it turns out that the eigenvalue density
depends strongly on the quark mass \cite{O,AOSV}. Hence we can not interpret the quark mass
as a test charge which, in a harmless manner, measures the chiral
condensate/electric field. 
To study the eigenvalue density we turn to the
microscopic scale of QCD \cite{SV} where the dimensionless combinations  
\be
\hm\equiv m\Sigma V \qquad {\rm and} \qquad \hmu^2\equiv \mu^2F_\pi^2V
\ee 
are kept fixed as the volume $V$ is taken to infinity ($\Sigma$ is the chiral
condensate and $F_\pi$ the pion decay constant). In this limit the
Compton wavelength of the pions is much larger than the linear extent of the 
volume and the zero momentum 
mode of the pions dominates the low energy effective
theory of QCD \cite{GL,LS} resulting in a partition function that is
given by a group integral.
Here we will access the microscopic limit of QCD by analyzing
the chiral random matrix model introduced previously \cite{O}. 
We will show that the unquenched eigenvalue density
is complex and oscillating when $\mu>m_\pi/2$ and that these oscillations are
responsible for the discontinuity of the chiral condensate \cite{OSV}. 
The complex eigenvalue density of the Dirac operator for QCD at
nonzero chemical potential was also studied for gauge fields given
by a liquid of instantons \cite{Schafer} and strong fluctuations in the
spectral density could be identified.


\section{The random matrix model}
\label{sec:rmt}

The random matrix partition function with $N_f$ quark flavors with mass $m$
and chemical potential $\mu$ is defined as \cite{O}
\be
 {\cal Z}_N^{N_f}(m;\mu) &\equiv& \int
 d\Phi d\Psi \ w_G(\Phi) w_G(\Psi)
 {\det}^{N_f}(\,{\cal D}(\mu) + m\,)  ,
\label{ZNfNb}
\ee
where the non-Hermitian Dirac operator is given by
\be
\label{dnew}
\mathcal{D}(\mu) = \left( \begin{array}{cc}
0 & i \Phi + \mu \Psi  \\
i \Phi^{\dagger} + \mu \Psi^{\dagger} & 0
\end{array} \right) ~.
\ee
The $(N+\nu)\times N$ matrices $\Phi$ and $\Psi$ are complex, $\nu$ is the
topological index, and $w_G$ is a Gaussian weight function  
\be
\label{wg}
w_G(X) ~=~ \exp( \, - \, N \, \tr \, X^{\dagger} X \, ) ~.
\ee
In this random matrix model, the microscopic limit is defined as the limit
$N\to \infty$ where
\be
\hat{m} = 2 N m  \ \ \ {\rm and} \ \ \ \hat{\mu}^2=2N \mu^2 
\ee    
are kept fixed. The random matrix partition function becomes identical to the
microscopic QCD partition function provided that we identify, see for example
\cite{AOSV}, 
\be\label{mapping}
\hat{m}= 2 N m & \to &  m \Sigma V, \\
\hat{\mu}^2 = 2 N \mu^2 & \to &  \mu^2 F_\pi^2 V. \nn
\ee
The main advantage of this matrix model as compared to the one originally
introduced in \cite{misha} is that an eigenvalue representation is known \cite{O}
\be
\label{epfnew}
{\cal Z}_N^{N_f}(m;\mu)  \sim
m^{\nu N_f } \int_{\mathbb{C}} \prod_{k=1}^{N} d^2z_k \,
{\cal P}^{N_f}(\{z_i\},\{z_i^*\}, m; \mu),
\ee
where the joint probability distribution reads
\be
\label{jpd}
{\cal P}^{N_f}(\{z_i\},\{z_i^*\},m;\mu) &=&  \frac{1}{\mu^{2N}} \left|\Delta_N(\{z_l^2\})\right|^2 \, 
\prod_{k=1}^{N} w(z_k,z_k^*;\mu) (m^2-z^2_k )^{N_f} .\nn \
\ee 
The Vandermonde determinant is defined as  
\be
\Delta_N(\{z^2_l\}) \equiv \prod_{i>j=1}^N (z_i^2-z_j^2),
\label{vander}
\ee
and the weight function includes a modified Bessel function,
\be 
w(z_k,z^*_k;\mu) &=& |z_k|^{2\nu+2} 
K_\nu \left( \frac{N (1+\mu^2)}{2 \mu^2} |z_k|^2 \right)
e^{-\frac{N (1-\mu^2)}{4 \mu^2}  
(z^2_k + \conj{z_k}^2)}. 
\label{wnew}
\ee
Given the eigenvalue representation we can employ the powerful method of
orthogonal polynomials in the complex plane developed in
\cite{FS,A03,AV,BI,BII,AP}.

\subsection{Orthogonal polynomials, partition function and chiral condensate}

The complex orthogonal polynomials corresponding to the weight function (\ref{wnew}) 
can be expressed in terms of complex Laguerre polynomials \cite{O}
\be
p_k(z;\mu) = \left( \frac{1-\mu^2}N\right )^k k! 
L_k^\nu \left ( -\frac{Nz^2}{1-\mu^2} \right).
\ee 
They satisfy the orthogonality relations
\be
\int_{\mathbb{C}}d^2z\ w(z,z^*;\mu)\ p_k(z;\mu)\ p_l(z;\mu)^* ~ 
 ~=~ \delta_{kl} ~ r_k^\nu ~,
\label{Jo1}
\ee
with the norm $r_k^\nu $ given by 
\be
\label{Norm}
r_k^\nu ~=~
\frac{  \pi \, \mu^2 ~ (1+\mu^2)^{2k+\nu} ~ k! ~ (k+\nu)!}
     {N^{2k +  2+\nu}}  ~.
\ee
The partition function for one fermion can be expressed in terms
orthogonal polynomials as
\be
\label{ZNf1}
Z_N^{N_f=1}(m;\mu) = m^\nu p_N(m;\mu).
\ee
Using this partition function (\ref{ZNf1}) we immediately find the chiral condensate
\be
\Sigma_N^{N_f=1}(m) = \frac{d}{dm}\log Z_N^{N_f=1}(m;\mu) = \frac{d p_N(m)/dm}{p_N(m)}+\frac{\nu}{m}.
\label{Sigma}
\ee
Since we have not yet taken the microscopic limit the chiral condensate
depends on $\mu$.

\section{The chiral condensate from the eigenvalue density using orthogonal polynomials}
\label{sec:OSVfromOP}

\begin{figure}[t]
  \unitlength0.9cm
  \begin{center}
  \begin{picture}(3.0,2.0)
  \put(-6.5,-11.){
  \psfig{file=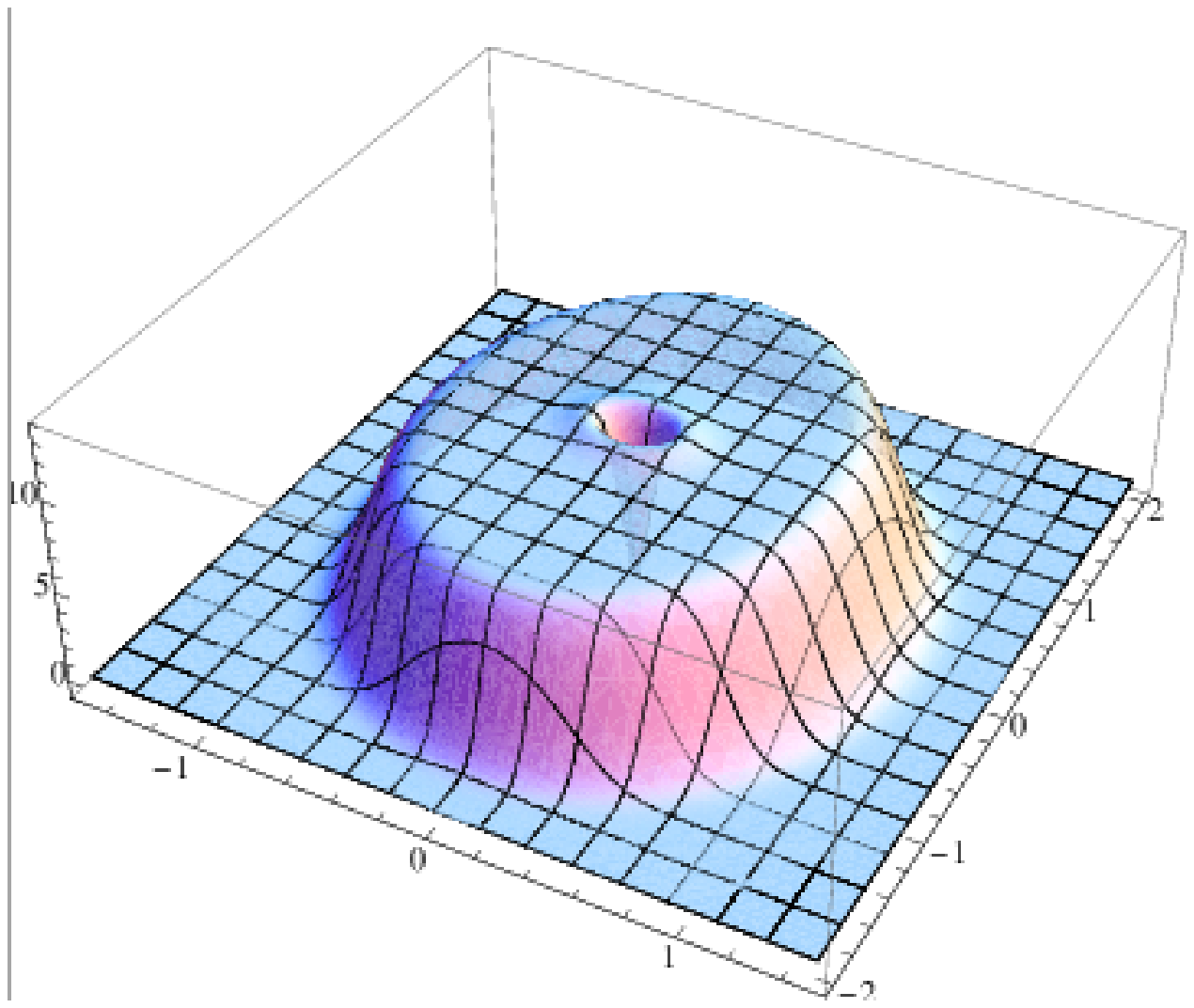,clip=,width=10.5cm}}
  \put(-1.5,-4.9){\bf\large Re$[z]$}
  \put(4.5,-2.0){\bf\large Im$[z]$}
  \put(-6.8,1.7){\bf \Large ${\rm Re}[\rho^{N_f=1}_{N}(z,z^*,m;\mu)]$}
  \put(-6.5,-20.){
  \psfig{file=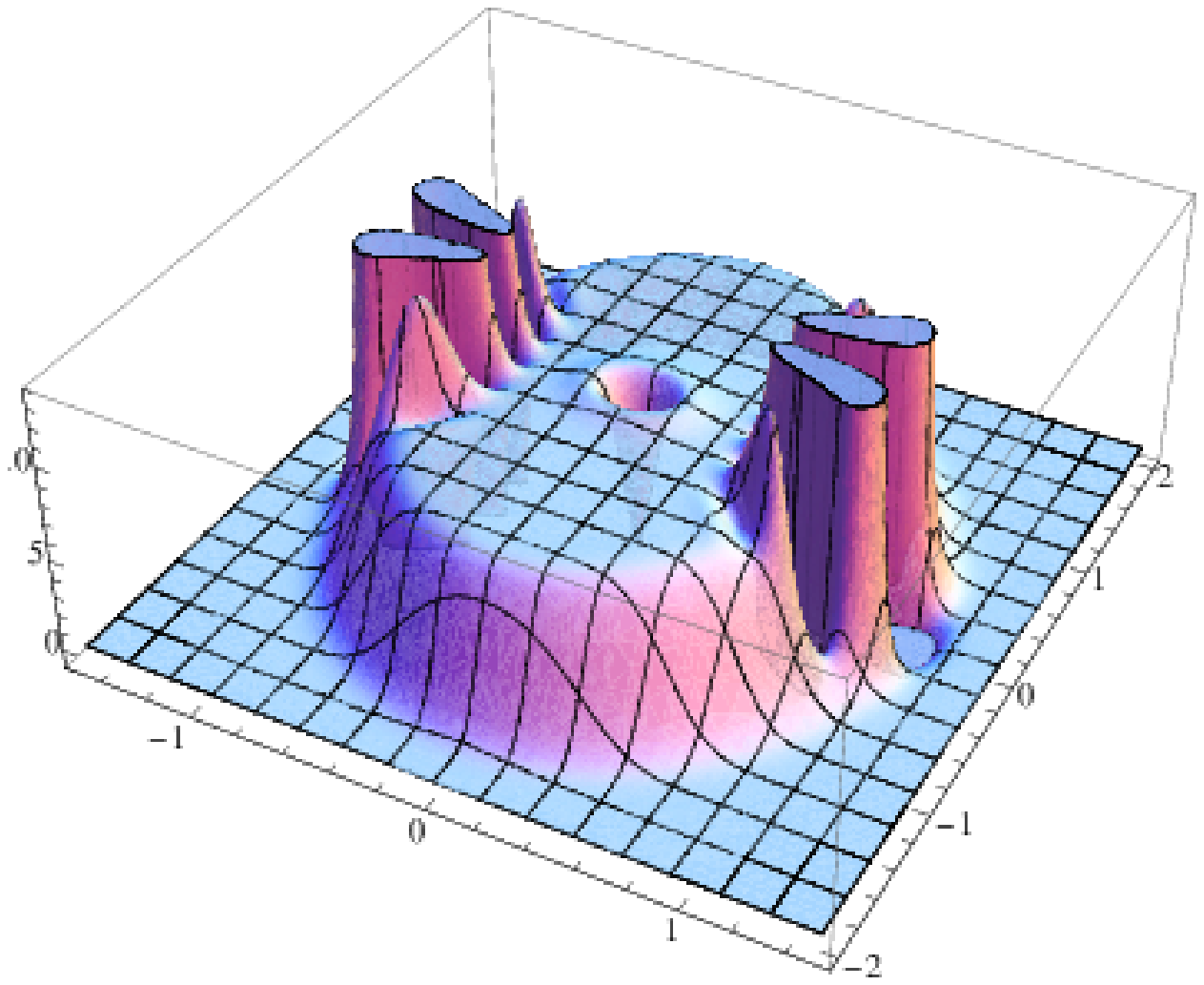,clip=,width=10.5cm}}
  \put(-1.3,-13.8){\bf\large Re$[z]$}
  \put(5,-10.4){\bf\large Im$[z]$}
  \put(-6.8,-7.){\bf \Large ${\rm Re}[\rho^{N_f=1}_{N}(z,z^*,m;\mu)]$}
  \end{picture}
  \vspace{12.5cm}
  \end{center}
\caption{\label{fig:rho} The eigenvalue density of the random matrix
  model (\ref{ZNfNb}) for $N=20$, $\mu=0.8$ and $\nu=0$.  
  {\bf Top:} The quark mass is well outside the support of the spectrum 
at $m=2.0$, and the
  eigenvalue density is real and positive.
  {\bf Bottom:} The quark mass is now inside the eigenvalue cloud at $m=0.6$,
  and  oscillations starting at $z=\pm m$ extend outward. Only the
  real part of the spectral density is shown -- 
the imaginary part is nonzero inside the oscillating region.} 
\end{figure}

The quenched eigenvalue density  given by
\be
\rho_N^{N_f=0}(z,z^*;\mu) = w(z,z^*;\mu)\sum_{k=0}^{N-1}\frac{p_k(z^*)p_k(z)}{r_k}.
\ee
is real and positive. The unquenched spectral density \cite{O} 
which can be expressed as
\be
\label{rhoNf1N}
\rho_N^{N_f=1}(z,z^*,m;\mu) =
w(z,z^*;\mu)\sum_{k=0}^{N-1}\frac{p_k(z^*)(p_k(z)-p_N(z)p_k(m)/p_N(m))}{r_k} 
\nn \\ \label{eq18}
\ee
is not invariant under complex conjugation. The asymmetric second term causes
complex valued oscillations, see figure \ref{fig:rho}.  

Here we will compute the chiral condensate (\ref{Sigma}) starting from the complex and
oscillating eigenvalue density given in (\ref{eq18}), 
\be
\Sigma_N^{N_f=1}(m) = \int_{\mathbb{C}} d^2z\frac{2m}{z^2-m^2} \rho_N^{N_f=1}(z,z^*,m;\mu).
\ee
We will show that all integrals can be carried out using the orthogonality
condition (\ref{Jo1}). The first step is to insert the above expression for
the unquenched density 
\be
\Sigma_N^{N_f=1}(m) = \int_{\mathbb{C}} d^2z\frac{2m w}{z^2-m^2}
\sum_{k=0}^{N-1}\frac{p_k(z^*)(p_k(z)-p_N(z)p_k(m)/p_N(m))}{r_k}.\;\;\;
\label{Sigma-raw}
\ee
We now use the fact that
\be
 \int_{\mathbb{C}} d^2z\frac{w(z,z^*;\mu) p_k(z^*)p_k(z)}{z^2-m^2} 
=\int_{\mathbb{C}} d^2z\frac{w(z,z^*;\mu) p_k(z^*)p_k(m)}{z^2-m^2},
\label{J1}
\ee
which is easily proved by first rewriting (the $c_k$'s are independent of $z$) 
\be
p_k(z)=(z^2-m^2)[p_{k-1}(z)+c_{k-2}p_{k-2}(z)+...]+p_k(m)
\label{J2}
\ee
and then using orthogonality, and find
\be
\Sigma_N^{N_f=1}(m) = \int_{\mathbb{C}} d^2z\frac{2m w}{z^2-m^2}\sum_{k=0}^{N-1}\frac{p_k(z^*)p_k(m)}{r_k}\frac{1}{p_N(m)}(p_N(m)-p_N(z)).\;\;\;
\ee
For $p_N(m)-p_N(z)$ we now insert (this is just Eq.~(\ref{J2}) rearranged a bit) 
\be
p_N(m)-p_N(z)=-(z^2-m^2)[p_{N-1}(z)+c_{j-2}p_{N-2}(z)+...]
\ee
and get
\be
\Sigma_N^{N_f=1}(m) = -\int_{\mathbb{C}} d^2z \ 2m
w\sum_{k=0}^{N-1}\frac{p_k(z^*)p_k(m)}{r_k p_N(m)}[p_{N-1}(z)+c_{N-2}p_{N-2}(z)+...].\nn\\
\ee
Using orthogonality we find (note that the $\mu$-dependent norms drop out)
\be
\Sigma_N^{N_f=1}(m) = \frac{2m}{p_N(m)}[p_{N-1}(m)+c_{N-2}p_{N-2}(m)+...].
\ee
Now, 
\be
\frac{d p_N(m)}{d m} 
  & = & \lim_{z\to m} \frac{p_N(z)-p_N(m)}{z-m}, \nn \\
  & = & \lim_{z\to m} 2m\frac{p_N(z)-p_N(m)}{z^2-m^2}, \nn \\
  & = & \lim_{z\to m} 2m [p_{N-1}(z)+c_{N-2}p_{N-2}(z)+...], \nn \\
  & = & 2m [p_{N-1}(m)+c_{N-2}p_{N-2}(m)+...],
\ee
so that 
\be
\Sigma_N^{N_f=1}(m) = \frac{ p_N'(m)}{p_N(m)}.
\ee
This is the desired relation (\ref{Sigma}) up to the term $\nu/m$
which is due to the contribution of $\nu$ exact zero modes that were
not included in the spectral density. The result of this section
originally appeared in \cite{OSV}. 

\vspace{3mm}

Note that the proof relies entirely on orthogonality properties of 
polynomials. The detailed cancellations leading to the simple form of the
chiral condensate are in this way linked to the orthogonality properties in
the complex plane. 

The above proof was given for finite $N$ and is of course also valid in the
universal microscopic limit. As shown in \cite{OSV}, the proof can also be
carried through in the microscopic limit without relying on the finite $N$ 
chiral random matrix model. That is, one can start from the microscopic limit
of the spectral density (\ref{rhoNf1N}), which is
\cite{O,AOSV}
\be\label{rhoNf1mikro}
\rho_{N_f=1}^{(\nu)}(\hat{z},\hat{z}^*,\hat{m};\hat\mu) 
& = & \frac{|\hat{z}|^2}{2\pi \hat\mu^2} 
K_\nu\left(\frac{|\hat{z}|^2}{4\hat\mu^2}\right)
\mbox{e}^{-\frac{\hat{z}^2+\hat{z}^{*\,2}}{8\hat\mu^2}}\\
&& \hspace{-3cm} \times\left(\int_0^1 dt \ t \ \mbox{e}^{-2\hat\mu^2 t^2}
I_\nu(\hat{z} t)I_\nu(\hat{z}^*
t)-\frac{I_\nu(\hat{z})}{I_\nu(\hat{m})}\int_0^1 dt \ t \
\mbox{e}^{-2\hat\mu^2 t^2}I_\nu(\hat{m} t)I_\nu(\hat{z}^* t)\right)\nn ,
\ee
and compute the chiral condensate using complex contour integration
techniques. In the microscopic limit the cancellations are even more dramatic
than at finite $N$. The
entire $\mu$ dependence of the eigenvalue density vanishes upon integration
and leaves us with a $\mu$ independent chiral condensate
\be
\Sigma(\hm)=\frac{I'_\nu(\hm)}{I_\nu(\hm)}.
\ee
Rather than repeating the
proof of \cite{OSV} we now focus on the Fourier components of
 the eigenvalue density
that are responsible for chiral symmetry breaking.

Note that the first term in (\ref{rhoNf1mikro}) is the
quenched eigenvalue density \cite{SplitVerb2} and the effect of unquenching 
is in the second term. This motivates the notation 
\be
\rho_{N_f=1}^{(\nu)} = \rho_Q^{(\nu)} - \rho_U^{(\nu)},
\label{qudec}
\ee
which we shall make use of below.

\section{The Fourier transform of the eigenvalue density}
\label{sec:rhoF}

To further expose the nature 
of the cancellations leading to a $\mu$ independent
chiral condensate  we now consider the Fourier transform of the
eigenvalue density. The motivation to do so is that the discontinuity in
the chiral condensate  is due to the
strongly oscillating part of the spectral density \cite{OSV}. 
To evaluate the chiral
condensate from the eigenvalue density  
\be
\Sigma_{N_f}(m) = \int dx dy \ \frac{\rho_{N_f}(x,y)}{x+iy-m}
\ee
illustrates the effects of the phase of the fermion determinant in
QCD at nonzero chemical potential. 
In particular for $\mu>m_\pi/2$ where the sign problem is acute \cite{sign},
it would be  desirable if we could find a real and positive 
reformulation of this integral.   

\vspace{3mm}

The Fourier transform of the eigenvalue density along
the imaginary axis is defined by
\be
\label{trhoDef}
\tilde{\rho}_{N_f}(x,t) & \equiv & \int_{-\infty}^\infty dy
e^{-iyt}\rho_{N_f}(x,y), \\
& = &  \left\langle \sum_k \delta(x-x_k)e^{-iy_k t} \right\rangle_{N_f} . \nn
\ee
Since we integrate over all $y$ the Fourier transform is necessarily
real.

The chiral condensate becomes
\be
\Sigma(m) 
& = & \int dx dy \frac{\rho(x,y)}{x+iy-m}, \nn \\
& = &\int dx dy \frac{1}{x+iy-m}\frac{1}{2\pi}\int dt e^{iyt} \trho(x,t),\nn  \\
& = &\frac{1}{2\pi} \int dx \int dt \trho(x,t) \int dy
\frac{e^{iyt}}{x+iy-m},\nn  \\
& = & \int dx \int dt \trho(x,t) e^{t(m-x)}\theta(x-m) .
\ee
Now the integrand is real and positive if $\trho$ is. Below we will see that
this is the case for the asymptotic limit of the unquenched microscopic density.

\subsection{The asymptotic limit of the microscopic spectral density}

In the limit where the microscopic variables $\hm$ and $\hmu$ are much larger
than unity, the expression for the eigenvalue density simplifies considerably.   
The quenched part of the microscopic eigenvalue distribution is simply constant
and equal to $1/4\hmu^2$ for Re$[\hat{z}]<2\hmu^2$. This leads to a chiral
condensate that  decreases linearly with the quark mass
\be
\Sigma_Q(\hm) = \frac{\hm}{2\hmu^2} \qquad {\rm for} \qquad |\hm| < 2\hmu^2 .
\label{SigmaQ}
\ee  
In the asymptotic limit the oscillating part of the microscopic eigenvalue density reads
\cite{OSV} (here and below $\hm\geq 0$)
\be
\rho_{U}(\hx,\hy) =
\frac{1}{4\pi\hmu^2}e^{(\hm-\hx-i\hy)(\hm+3\hx-8\hmu^2-i\hy)/(8\hmu^2)} . 
\ee
Obviously the amplitude grows exponentially with the volume whereas the
period is inversely proportional to the volume.

\begin{figure}[t]
  \unitlength0.90cm
  \begin{center}
  \begin{picture}(3.0,2.0)
  \put(-6.5,-11.){
  \psfig{file=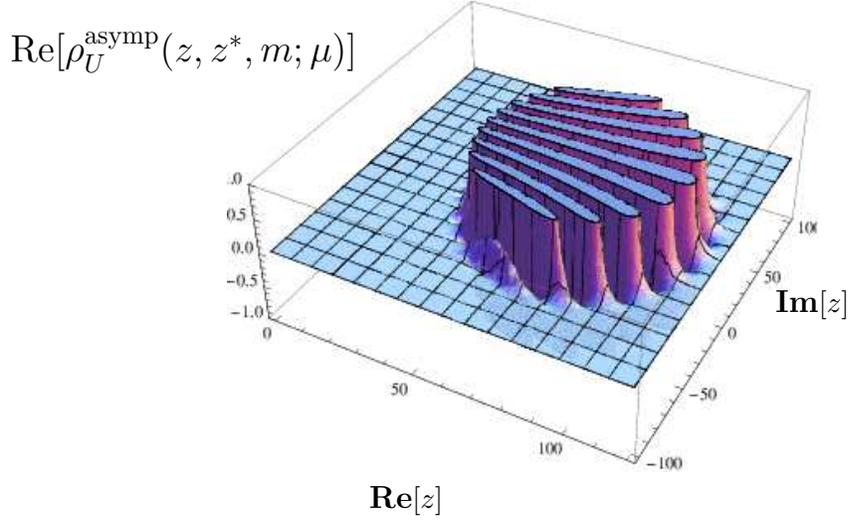,clip=,width=10.5cm}}
  \put(-1.5,-4.9){\bf\large Re$[z]$}
  \put(4.5,-2.0){\bf\large Im$[z]$}
  \put(-6.8,1.7){\bf \Large ${\rm Re}[\rho^{\rm asymp}_{U}(z,z^*,m;\mu)]$}
  \put(-6.5,-20.){
  \psfig{file=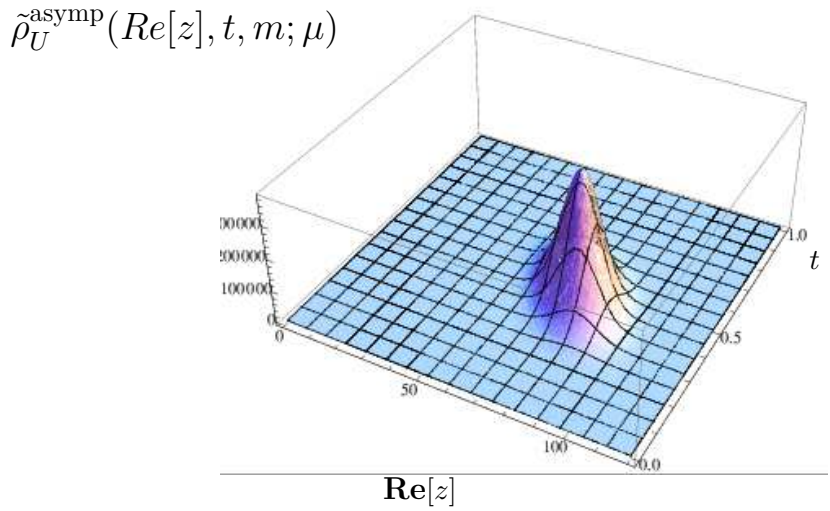,clip=,width=10.5cm}}
  \put(-1.3,-13.8){\bf\large Re$[z]$}
  \put(5,-10.4){\bf\large $t$}
  \put(-6.8,-7.){\bf \Large $\trho^{\rm asymp}_{U}(Re[z],t,m;\mu)$}
  \end{picture}
  \vspace{12.7cm}
  \end{center}
\caption{\label{fig:rhoUasymp} {\bf Top:} The asymptotic limit of the
  oscillating  part of the unquenched eigenvalue density. {\bf Bottom:} The
  Fourier transform of the asymptotic limit of the oscillating part of the
  unquenched eigenvalue density. In both cases $\hm=30$ and $\hmu=7$.}
\end{figure}

By simple Gaussian integration it follows that the Fourier transform of
$\rho_U$ is    
\be
\trho_U(\hx,\htt) =
\frac{1}{\sqrt{2\pi\hmu^2}}
e^{-2\hmu^2(\htt-1)^2-\hx^2/2\hmu^2-\hm\htt-(\htt-2)\hx +2\hm}.
\ee
It is a positive definite and nicely behaved function with maximum
at $\htt=1-(\hm+\hx)/4\hmu^2$ (see Fig. \ref{fig:rhoUasymp}).
The contribution to the chiral condensate of
the unquenched part of the spectral density thus
 becomes an integral over a positive definite function
of the real part of the eigenvalue, $x$, and the Fourier coordinate, $t$,
\be
\Sigma_U(\hm) 
& = & \frac{1}{\sqrt{2\pi\hmu^2}} \int_{-2\hmu^2}^{2\hmu^2} d\hx
\int_{-\infty}^\infty d\htt 
e^{-2\hmu^2(\htt-1)^2-\hx^2/2\hmu^2-2(\htt-1)\hx} \theta(\hx-\hm) .  \;\;\;
\ee
A simple computation yields
\be
\Sigma_U(\hm) 
&= & \frac{1}{\sqrt{2\pi\hmu^2}} \int_{-2\hmu^2}^{2\hmu^2} d\hx\theta(\hx-\hm)
e^{-\hx^2/2\hmu^2}
\int_{-\infty}^\infty d\hat{q} e^{-2\hmu^2\hat{q}^2-2\hat{q}\hx}, \nn \\
&= &\frac{1}{\sqrt{2\pi\hmu^2}} \int_{-2\hmu^2}^{2\hmu^2} d\hx\theta(\hx-\hm)
e^{-\hx^2/2\hmu^2}
\sqrt{\frac{\pi}{2\hmu^2}} e^{\hx^2/2\hmu^2}, \nn \\
&=& \frac{1}{2\hmu^2}\int_{-2\hmu^2}^{2\hmu^2} d\hx\theta(\hx-\hm), \nn \\
& = & 1-\frac{\hm}{2\hmu^2}.
\ee
The full chiral condensate is obtained by adding the quenched part
(\ref{SigmaQ}), note that the $\mu$ dependent term drops out. The message to
take away from this derivation  is that the calculation of the chiral
condensate may simplify by using the
mixed Fourier transform of
the spectral density introduced above. 



\section{Conclusions}

Chiral symmetry breaking has been linked to complex 
oscillations on the microscopic
scale. The cancellation of these violent oscillations, which result in the
correct value of the chiral condensate,
demonstrates the
numerical difficulties in dealing with the sign problem when the quark mass
is inside the support of the eigenvalues. At low temperatures this occurs for
$\mu>m_\pi/2$.  We have shown that the mechanism of these cancellations
can be understood in terms of orthogonality relations for the orthogonal
polynomials corresponding to a random matrix theory that describes 
the microscopic limit of QCD. The derivation is even valid for finite
size random matrices. 
The fact that the oscillations predominantly occur parallel to
the imaginary axis motivated us to study the Fourier transform of the
eigenvalue density in the imaginary part of the eigenvalue. In the asymptotic
limit the resulting Fourier transformed eigenvalue density is real and
positive, and in this formulation a probabilistic interpretation of the
contributions to the chiral condensate is possible. 

It would be most interesting to generalize these results to the 
$p$-domain of chiral perturbation theory. 

\section*{Acknowledgments}
It is a pleasure to thank the FTPI faculty and staff for an
exciting workshop and Poul Henrik Damgaard for stimulating discussions. 
This work was supported  by U.S. DOE Grant
No. DE-FG-88ER40388 (JV), the Villum
Kann Rassmussen Foundation (JV), the Danish National Bank (JV) and the
Danish Natural Science Research Council (KS).

\bibliographystyle{ws-procs9x6}
\bibliography{ws-pro-sample}

\end{document}